\documentclass{aa}

\usepackage{graphicx}
\usepackage{txfonts}
\usepackage[normalem]{ulem}
\usepackage{xcolor}
\usepackage{booktabs}
\begin{document}

%   \title{The Effect of Metallicity on the Capability of Detecting the Rotation Periods of Solar‐like Stars}

% suggestion
\title{Effect of metallicity on the detectability of rotational periods in solar-like stars}
    
 %  \subtitle{I. maybe }

   \author{V. Witzke\inst{1}
          \and
          T. Reinhold\inst{1}
          \and
          A. I.\ Shapiro\inst{1}%\fnmsep\thanks{Just to show the usage
        %   of the elements in the author field}
          \and
          N. A.\ Krivova\inst{1}
          \and
          S. K.\ Solanki\inst{1,2}
          }

   \institute{Max Planck Institute for Solar System Research, Justus-von-Liebig-Weg 3, 37077 G\"ottingen, Germany\\
              \email{witzke@mps.mpg.de}
         \and
         School of Space Research, Kyung Hee University, Yongin, Gyeonggi, 446-701, Republic of Korea       
             }

%   \date{\today}

% \abstract{}{}{}{}{} 
% 5 {} token are mandatory
 
\abstract{Accurate determination of stellar rotation periods is important for estimating stellar ages as well as for understanding stellar activity and evolution. While for about thirty thousand stars in the \textit{Kepler} field rotation periods can be determined, there are over hundred thousand stars, especially with low photometric variability and irregular pattern of variations, for which rotational periods are unknown. %Here, we investigate the link between the detectability of rotational periods and the metallicity for solar-like stars. 
Here, we investigate the effect of metallicity on the detectability of rotation periods. This is done by synthesising light curves of hypothetical stars, which are identical to our Sun, with the exception of the metallicity. These light curves are then used as an input to the period determination algorithms.
%Using a stellar brightness variations model based on the well established model of solar brightness variations, SATIRE, we investigate the effect %of metallicity on the detectability of rotational periods. 
We find that the success rate for recovering the rotation signal has a minimum close to the solar metallicity value. This can be explained by the compensation effect of facular and spot contributions. In addition, selecting solar-like stars with near-solar effective temperature, near solar photometric variability,  and with metallicity between M/H = -0.35 and M/H = 0.35 from the \textit{Kepler} sample, we analyse the  fraction of stars for which rotational periods have been detected as a function of metallicity. In agreement with our theoretical estimate we found a local minimum for the detection fraction close to the solar metallicity. We further report rotation periods of 87~solar-like \textit{Kepler} stars for the first time.
}


\maketitle

%% ------------------------------
%%%%%%%%%%%%%%%%%%%%%%%%%%%%%%%%%%%%%%%%%%%%%%%%%%%%%%%%%%%%%%%%%%%%%%%%%%%%%%%%%%%%%%%%%%%%%%%%%%%%%%%%%%%%%%%%%
% I N T R O D U C T I O N  
%%%%%%%%%%%%%%%%%%%%%%%%%%%%%%%%%%%%%%%%%%%%%%%%%%%%%%%%%%%%%%%%%%%%%%%%%%%%%%%%%%%%%%%%%%%%%%%%%%%%%%%%%%%%%%%%%
%-------------------------------------------------------------------

\section{Introduction}
% why rotational periods matter
The rotation period of a star is a  fundamental stellar parameter that is closely linked to the stellar age and determines its magnetic activity.
\citet{1972ApJ...171..565S} was first to demonstrate that the equatorial rotational velocity and magnetic activity (the latter expressed via the emission in the \ion{Ca}{ii} H and K line cores)  decrease with stellar age $t$ as $1/\sqrt{t}$. Thereafter, an age-rotation-activity relation was established that indicates that young stars generally rotate faster and are more active compared to old stars \citep{Noyes1984}. Hence, determining stellar rotation periods is crucial for assessing the stellar age \citep{1989ApJ...343L..65K, 2003ApJ...586..464B}, which otherwise can be, for example measured through asteroseismology. 
%cannot be measured directly.

%
% previous studies: Kepler
The rotational signal is imprinted in the stellar photometric variability, which  is caused by transits of magnetic features (such as spots and faculae) across the visible disc and by their temporal evolution. With the advent of large photometric surveys, such as that by the \textit{Kepler} mission, stellar photometric light curves have been measured for an extensive number of main-sequence stars. This allowed the determination of rotational periods for about thirty thousand stars \citep{McQuillan2013a, McQuillan2013b, Reinhold2013, Walkowicz2013, Nielsen2013, McQuillan2014, doNascimento2014, Garcia2014, Reinhold2015, Ceillier2016, Buzasi2016}.
Among these studies, the largest homogeneous set of rotation periods was derived by \citet{McQuillan2014} measuring rotation periods for 34,030 \textit{Kepler} stars. Despite this huge number of stars with determined rotation period, a significant rotation signal could not be detected  in an even bigger sample of 99,000 stars (see Table~2 in \citealt{McQuillan2014}). 
Stellar samples with known rotation periods form the basis for many studies, ranging from  galactic evolution to the solar-stellar connection \citep{Buzasi2016, Davenport2017, Davenport2018, Reinhold2019, Reinhold_sub, Notsu2019, vanSaders2019}. Consequently, the conclusions drawn in these studies might be strongly biased towards the behaviour of stars for which rotation periods could be determined.

% observational biases
To investigate potential detectability biases, it is  essential to understand possible physical reasons for a lack of the rotational signal in these stars. Our Sun provides a good example of a star whose highly irregular temporal profile of the photometric variability hampers the determination of the rotational period.
It has been shown that the irregular profile of solar variability is associated with the cancellation of facular and spot rotational signals in broad-band photometric measurements \citep{2017NatAs...1..612S,paperI_Eliana_Sasha,Reinhold2019}. 
One can expect that such a cancellation might also apply to other solar-like stars, in particular, for those with near-solar level of magnetic activity. This assumption is in line with recent results that the low success rate of less than 20\%  for the period determination in stars with near-solar effective temperatures  (see Fig.~14 in \citealt{vanSaders2019}) cannot be explained by the low signal-to-noise ratio in their photometric records.  % low photometric variability amplitude. This might explain the small number of solar-like stars for which rotation periods are known \citep{vanSaders2019}.
In addition, \citet{Reinhold2019} proposed that the cancellation of bright faculae and dark spots leads to a non-detection at intermediate rotation periods in the \textit{Kepler} field.

% aim of the paper
It was shown earlier  that  metallicity has a significant effect on facular contrasts, and thus on stellar variability in solar-like stars \citep{0004-637X-852-1-46, 2018A&A...619A.146W}.  In this work, our aim is to understand the effect of metallicity on the detectability of the rotation periods in solar-like stars.
We employ an extended version of the widely used model of solar brightness variability, the SATIRE-S (Spectral And Total Irradiance REconstruction, \citealt{2000A&A...353..380F, 2003A&A...399L...1K}). 
This model reconstructs stellar brightness variability based on solar magnetograms and disc images. To study the influence of metallicity on the detectability of rotational periods we generate light curves for different metallicities, and calculate the success rates for recovering the rotational signal by using periodograms.
Subsequently,  we analyse a sample of \textit{Kepler} stars with near-solar values of effective temperature,  photometric variability, and rotation periods, and compare the  effect of metallicity on the detectability of rotational periods in observations and the theoretical model.

% ------------------------------
%%%%%%%%%%%%%%%%%%%%%%%%%%%%%%%%%%%%%%%%%%%%%%%%%%%%%%%%%%%%%%%%%%%%%%%%%%%%%%%%%%%%%%%%%%%%%%%%%%%%%%%%%%%%%%%%%
% R E S U L T S 
%%%%%%%%%%%%%%%%%%%%%%%%%%%%%%%%%%%%%%%%%%%%%%%%%%%%%%%%%%%%%%%%%%%%%%%%%%%%%%%%%%%%%%%%%%%%%%%%%%%%%%
\section{Rotation period detectability in solar-like stars}

%%%%%%%%%%%%%%%%%%%%%%%%%%%%%%%%%%%%%%%%%%%%%%%%%%%%%%%%%%%%%%%%%%%%%%%%%%%%%%%%%%%%%%%%%%%%%%%%%%%%%%%%%%%%%%%%%
%------------------------ using the model :
%%%%%%%%%%%%%%%%%%%%%%%%%%%%%%%%%%%%%%%%%%%%%%%%%%%%%%%%%%%%%%%%%%%%%%%%%%%%%%%%%%%%%%%%%%%%%%%%%%%%%%%%%%%%%%%%%
%
%
\begin{figure}
%   \centering
    \resizebox{\hsize}{!}{\includegraphics{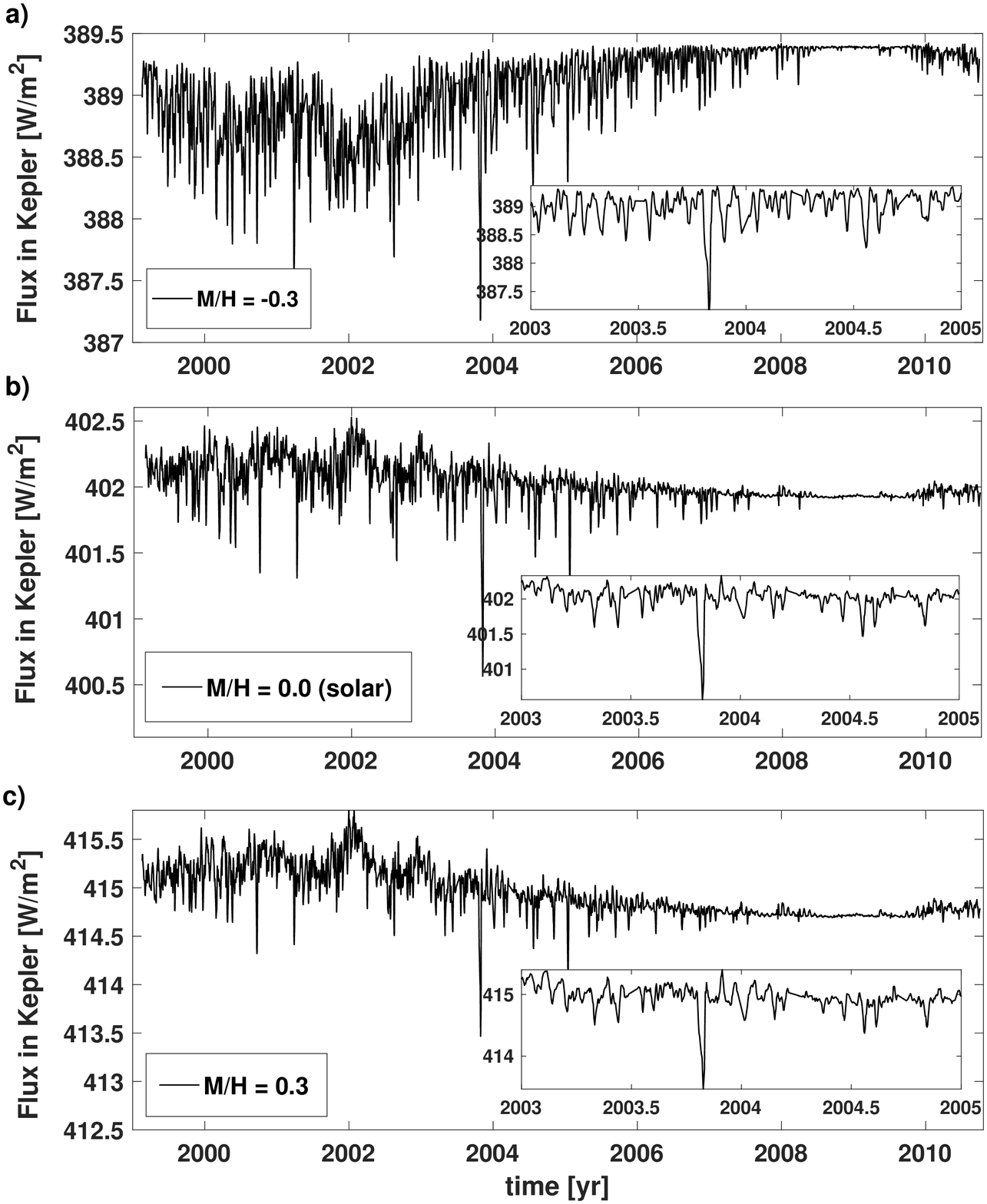}}
    \caption{Model light curves for different M/H values in the \textit{Kepler} pass band. a) M/H = -0.3, b) solar value, c) M/H = 0.3. The insets show shorter time intervals to better visualise the rotational variability.}
    \label{fig:0201}
\end{figure}

\subsection{Modelling}
\label{subsec:Modelling}
In this section we study the effect of metallicity on the success rate of  determining  rotational periods in solar-like stars. % $\rm R_{var} < 0.18\%$, 
 For that, we employ the SATIRE \citep[][]{2003A&A...399L...1K} model, which attributes stellar variability to the time-dependent contributions from magnetic features on the stellar surface, which can be divided into dark (spots and pores) and bright (faculae and network) features. 
Recently, this  model  was generalised for calculating variability of stars with  different metallicities \citep{2018A&A...619A.146W} by re-calculating spectra of the quiet and magnetic stellar regions.  Here we follow up this approach and generate light curves in the  \textit{Kepler} filter for stars identical to the Sun in all aspects except  metallicity. In particular, we  assumed the  solar distribution of magnetic features but recalculated their contrasts as a function of metallicity. 
%with solar distribution of magnetic features but different metallicities.
%use the SATIRE model together with emergent spectra for different metallicities to generate light curves as observed in the \textit{Kepler}-passband.  A more detailed description of the model can be found in \citet{2003A&A...399L...1K, 2014A&A...569A..38S, 2018A&A...619A.146W}. 

Figure~\ref{fig:0201} shows example light curves for three different metallicity values obtained by using the solar  magnetic feature distributions observed during solar cycle 23, which was a cycle of intermediate strength. Since the time series of  magnetic feature distribution is derived from SoHO/MDI observations \citep{1995SoPh..162..129S}, the modelled light curves are as seen by an observer on Earth.
%the same apparent position as viewed from Earth
For  $\rm M/H = -0.3$ (Fig.~\ref{fig:0201}a), the flux decreases during the activity cycle maximum in 2001--2003, indicating a spot-dominated variability on the time-scale of the 11-year activity cycle. The opposite is observed for the solar case (Fig.~\ref{fig:0201}b), and a star with $\rm M/H = 0.3$ (Fig.~\ref{fig:0201}c), where the flux increases during the activity maximum. These three cases  demonstrate the transition from spot-dominated   to faculae-dominated regimes on the activity cycle time-scale. 
%\cs{Do you want to explain here and mention dark faculae? - Actually I think it fits better when we explain the power spectra from faculae, spot and total light curves... }
% 

We measure the photometric variability of a star on the rotational time-scale by using the variability range $\rm R_{var}$. This quantity is defined  as the difference between the 95th and 5th percentile of the sorted differential intensities in the light curve \citep{Basri2010, Basri2011}. 
Table~\ref{table:1} illustrates how the amplitude of $\rm R_{var}$ depends on the metallicity.
The $\rm R_{var}$  values in Table~\ref{table:1} are obtained by  splitting the  1998--2011  light curves in subsequent 90-day intervals, 
and calculating the median of all individual $\rm R_{var}$ values for each of the intervals.
%

%$\rm R_{var}$ is defined  as the difference between the 95th and 5th percentile of the sorted differential fluxes in the light curve \citep{Basri2010, Basri2011}. }}

We find a slight  increase in  $\rm R_{var}$ for $\rm M/H = 0.3$ ($\rm R_{var} =0.074$ compared to $ \rm R_{var} = 0.062$ for the solar case) and a significant increase for $\rm M/H = -0.3$, $\rm R_{var} = 0.125$ (see Table~\ref{table:1}). This behaviour in $\rm R_{var}$ is caused by the effect of metallicity on the balance of spot to facular contrast. It is similar to the effect found in \citet{2018A&A...619A.146W}, but not as pronounced because the variability on the rotation time-scale, $\rm R_{var}$,  is mainly attributed to spots, whereas the predominant contribution to the variability on the magnetic activity time-scale for the Sun comes from faculae. We note that our approach employs 1D atmosphere models, which do not take 3D effects into account. While 3D effects  might lead to slightly different quantitative results, the overall trends are well captured in 1D.

%How the changed metallicity affects the detectability will be discussed in more detail below.

\begin{figure}
%   \centering
    \resizebox{\hsize}{!}{\includegraphics{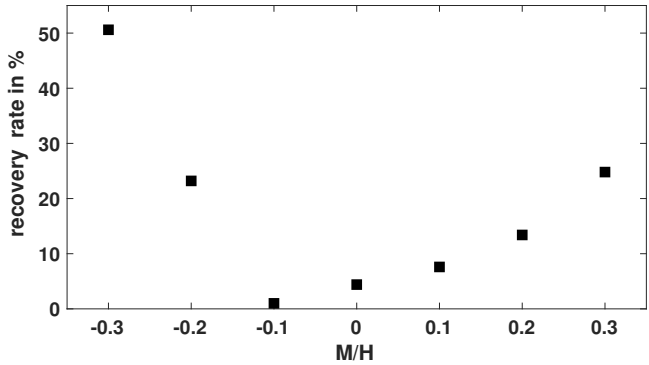}}
    \caption{Recovery rates for determining the rotational period in the range of 24--30 days for different metallicities.}
    \label{fig:0202}
\end{figure}

%This asymmetric behaviour is possibly caused by change in spot and facular contrasts with metallicity, which is discussed in more detail below.\\ 
%
%  
\begin{table}
\setlength\tabcolsep{5pt}
\renewcommand{\arraystretch}{1.5}
\caption{$\rm R_{var}$ with metallicity. $\rm R_{var}$ is given in percent for all metallicity values that we investigate using the SATIRE model. }              
\label{table:1}      
\centering                                     
\begin{tabular}{|c | c| c| c| c| c| c| c| }         
\hline\hline                       
M/H        & -0.3 & -0.2 & -0.1 & 0.0 & 0.1 & 0.2 & 0.3  \\   
\hline\hline                                 
 $\rm R_{var}$ &  0.125  &  0.070 & 0.062 &  0.062 & 0.064 &  0.068 & 0.074  \\
\hline\hline                                             
\end{tabular}
\end{table}
We investigate whether the detection of the Carrington synodic rotation, i.e.~the rotation signal of 27.3 days, is possible for such stars.
In order to emulate light curves of {\textit Kepler} stars, we randomly chose four-years intervals in the generated light curves. As we expect the detectability to depend on the phase of the activity cycle, we aim to capture the  full activity cycle 23 (mid 1996 --  end 2008). For that,  the 500 randomly chosen four-year intervals  were restricted to lie within mid 1994 and end of 2011.  The additional two years before and after the cycle 23 are needed to obtain a uniform distribution of the  positions of the middle point  of the four-year intervals within the cycle 23. 
To obtain the recovery rate of the solar rotation period,  we applied generalised Lomb-Scargle periodograms \citep{Zechmeister2009} to each realisation of the four-year intervals, and searched for a rotational signal. First, we determined the standard deviation, $\sigma$, of the power in the range of 0 to 50 days in the periodogram.  Only if the highest power peak in the periodogram satisfied the following four criteria, we assumed it as a successfully recovered  rotation period: i) it is the highest peak in the range of 0 to 50 days, ii) the peak is higher than $5 \sigma$, iii)  the second highest peak is at least 20\% lower, and iv) the  rotational period associated with this peak is between 24 and 30 days. Figure~\ref{fig:0202} shows the recovery rates in percent for different metallicities.
It shows that for stars with near solar metallicity the detection rate is very low. This is because of the absence of a significant peak in the periodogram for almost all of the 500 randomly chosen time intervals for such metallicities.  The detection rate increases significantly for metallicity values that are about 0.2 dex different from the solar one, reaching more than 55\% for M/H = -0.3.
%
%   

\begin{figure}
   
\resizebox{\hsize}{!}{\includegraphics{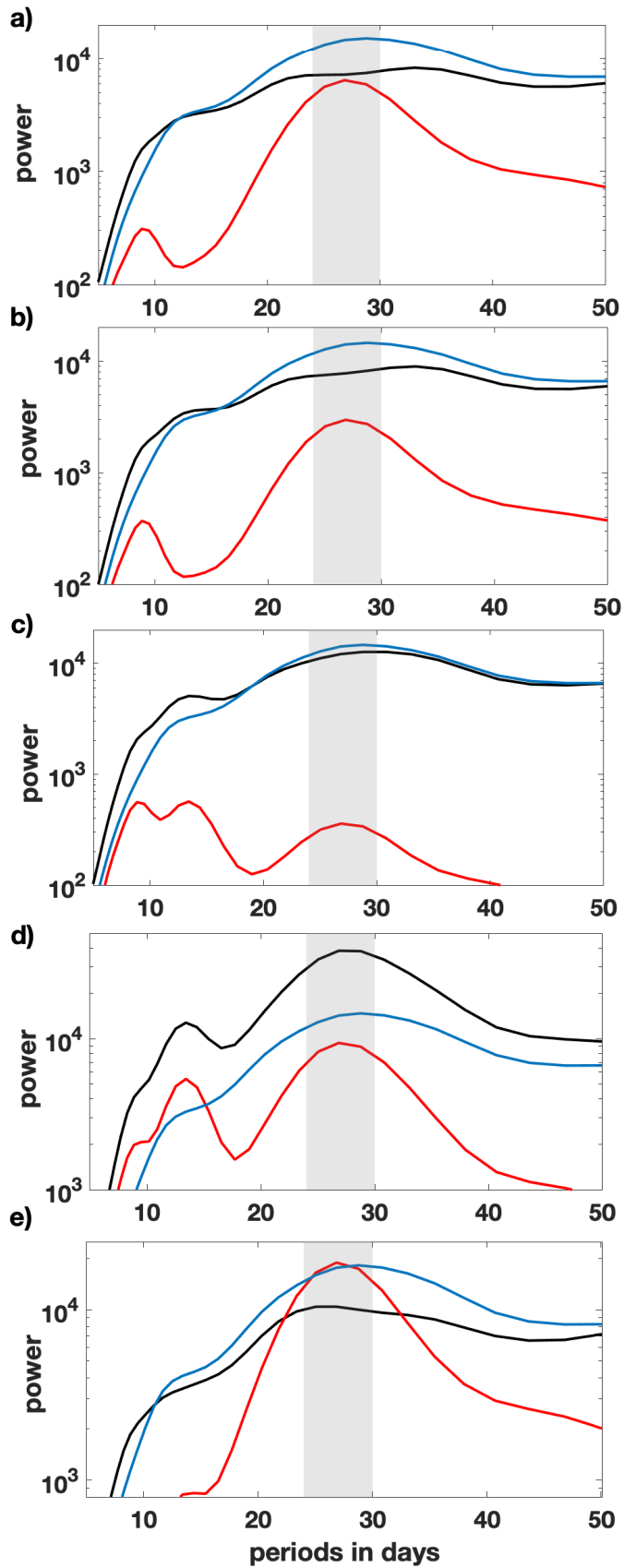}}
       \caption{power spectra of the facular- (red), spot (blue) and total (black) components for different metallicities. a) The solar case as reference, b) $\rm M/H = -0.1 $, c) $\rm M/H = -0.2$,  d) $\rm M/H = -0.3 $,  e)   $\rm M/H = 0.3 $. The grey area marks periods of 24--30~days. }
       \label{fig:0203}
\end{figure}

%%%%%%%%%%%% 
To understand the dependence of the success rates on the metallicity, we investigate the facular and spot components of the light curves separately. The SATIRE model uses decomposed spectral fluxes from quiet regions, facular regions, and spot regions, which enables us to generate separate light curves for the facular and spot components. 
We studied the contribution of the facular- and spot- components to the rotational signal. For a better illustration we  employed global wavelet power spectra instead of periodograms.  We considered the time interval of the whole cycle 23, and we calculated the power spectra using Morlet wavelets of the order six.
In Fig.~\ref{fig:0203} the power spectra for the cases with $\rm M/H = -0.3$, $\rm M/H =-0.2$,  $\rm M/H =-0.1$, $\rm M/H = 0.0$ and $\rm M/H=0.3$  are shown, where the solar case is included as a reference. 

Since we are interested in the solar rotational signal, we focus on the range of 24--30~days. For the solar case (Fig.~\ref{fig:0203}a), the  power obtained from the total light curve lies in between the power of the facular and the spot components. Although both the facular- and the spot power spectra show a peak, the  contribution of the bright faculae  cancels  the brightness contribution of the dark spots \citep{2017NatAs...1..612S}. Thus, there is no pronounced signal in the total power spectrum on the rotational time-scale. Since the success rates above were calculated for 500 different intervals, some of the intervals correspond to the activity minimum, where no or only few spots occur. For such intervals, it is possible to find the rotational signal due to the faculae \citep{Lanza_Shkolnik2014}.
The case with $\rm M/H = -0.1$ has a similarly flat power spectrum around the 24--30~days interval. The power of the total light curve lies in between the power of the facular and the spot components, but the power of the facular component is lower compared to the $\rm M/H =0.0$ case. This difference affects the compensation between faculae and spots, and thus can explain the lower success rates for $\rm M/H =-0.1 $ (see Fig.~\ref{fig:0202}).

For the case with $\rm M/H = -0.2$ the power of the total light curve lies also between the power of the facular and spot component, but the power of  the facular component is very small. The decreased power in the faculae results from a smaller facular contrast at the limb, and even a slightly darker faculae at the disc centre (see Fig.~\ref{fig:app02}). Thus the spot darkening prevails, and a determination of the rotational period becomes possible. Decreasing the metallicity further leads to a significant darkening of faculae at the disc centre (see Fig.~\ref{fig:app02}). Such that the facular contribution adds onto the spot contribution,  which becomes evident from Fig.~\ref{fig:0203}d), where the total component shows a clear peak and is above the facular and the spot components.  In contrast to this, at large metallicity values, e.g.~$\rm M/H = 0.3$  facular brightening becomes dominant (see Fig.~\ref{fig:0203}e). This also results in an increased detectability of the rotational period.

%%%%%%%%%%%
%
%
\begin{figure}
    \resizebox{\hsize}{!}{\includegraphics{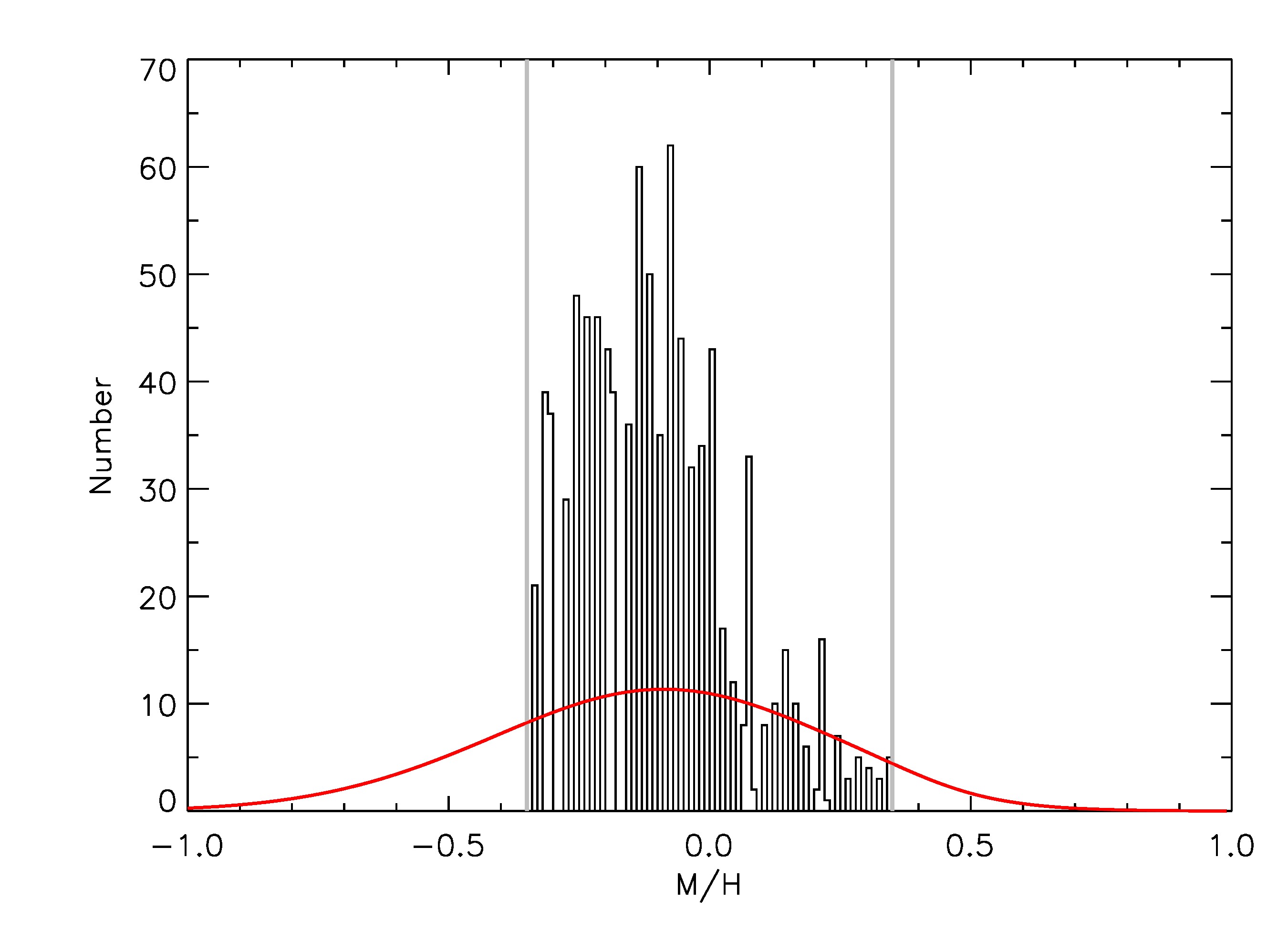}}
    \caption{Metallicity distribution of all selected stars.  The sample contains 911 stars with $\rm R_{var} < 0.18\%$. Black: histogram of stars in the metallicity range $\rm -0.35 < M/H < 0.35$, which is indicated by the grey lines. Red: The metallicity distribution using normalised Gaussian functions to account for the metallicity measurement uncertainty.}
    \label{fig:0100}
\end{figure}

\begin{figure}
  \includegraphics[width=8.5cm]{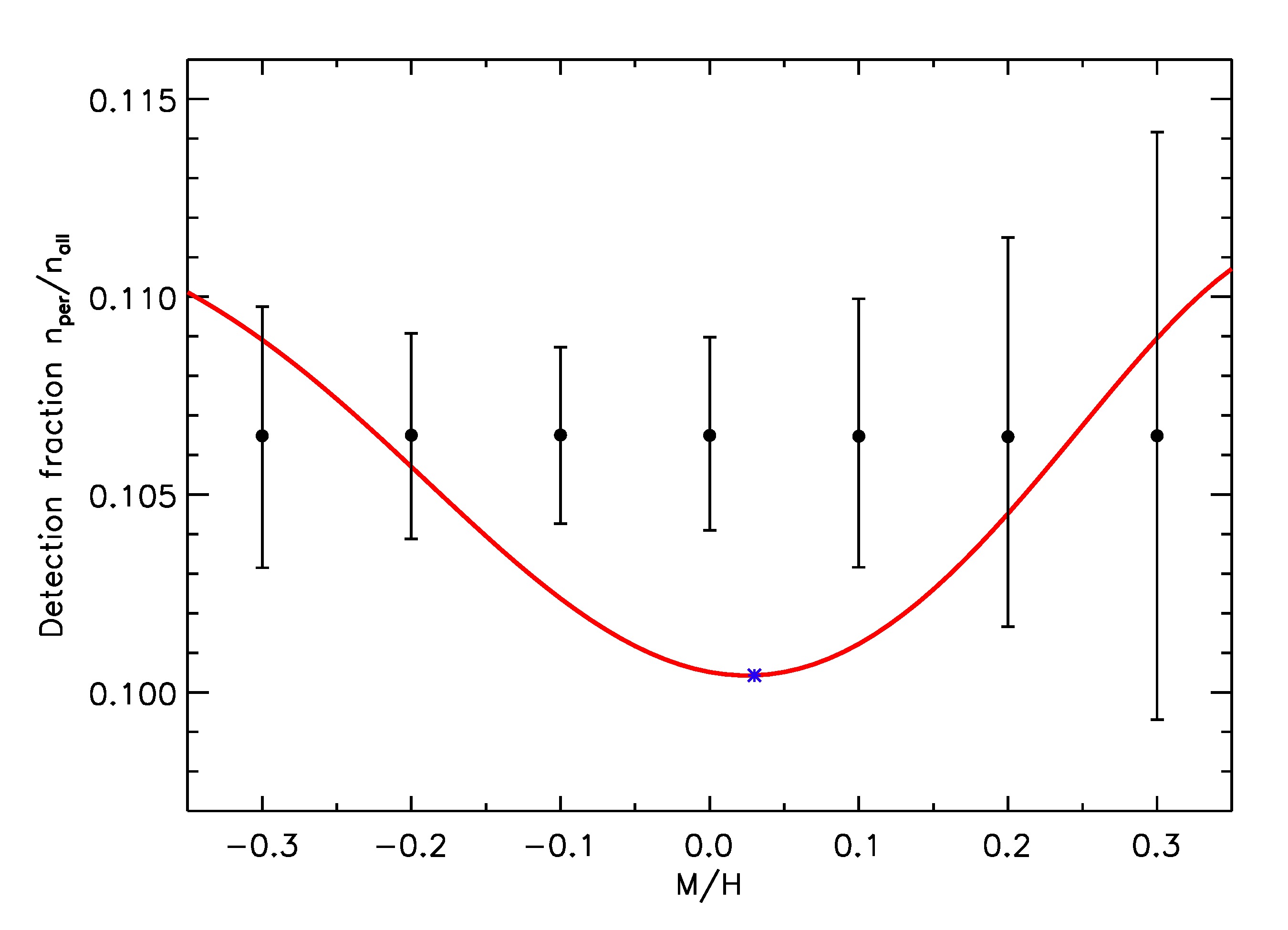}
  \caption{ Detection fraction as a function of metallicity (red line). The detection fraction is defined as the ratio of the density distribution of stars with known rotation periods, $\rm n_{per}$, to the density distribution of the total number of stars, $n_{all}$.   The black dots indicate the expected ratio, which is derived as the ratio of all periodic stars to all stars in a sample. The error bars correspond to the standard deviation using 10,000 realisations by randomly choosing 97 stars. The blue star indices the minimum of the detection fraction.}
  \label{fig:0101}
\end{figure}
%
%%%%%%%%%%%%%%%%%%%%%%%%%%%%%%%%%%%%%%%%%%%%%%%%%%%%%%%%%%%%%%%%%%%%%%%%%%%%%%%%%%%%%%%%%%%%%%%%%%%%%%%%%%%%%%%%%
% Observations 
%%%%%%%%%%%%%%%%%%%%%%%%%%%%%%%%%%%%%%%%%%%%%%%%%%%%%%%%%%%%%%%%%%%%%%%%%%%%%%%%%%%%%%%%%%%%%%%%%%%%%%%%%%%%%%%%%
%%%%%%%%%%%%%%%%%%%%%%%%%%%%%%%%%%%%%%%%%%%%%%%%%%%%%%%%%%%%%%%%%%%%%%%%%%%%%%%%%%%%%%%%%%%%%%%%%%%%%%%%%%%%%%%%%
\subsection{Analysis of \textit{Kepler} observations }
We now investigate whether a similar trend with metallicity can also be found for solar-like \textit{Kepler} stars.
%in observations, we study the dependence of the rotation period detectability on the  metallicity for solar-like \textit{Kepler} stars.
We follow the selection criteria of \citet{Reinhold_sub} to focus on solar-like main-sequence stars. In particular, we select a sample of stars with effective temperatures in the range 5600--5900\,K (which corresponds to the solar effective temperature $\pm 150$\,K), surface gravities $\log\,g>4.2$, and metallicities in the range $-0.35 < \rm M/H < 0.35$. Stars fainter than 15th \textit{Kepler} magnitude are discarded. Stellar fundamental parameters ($\rm T_{eff}$, log\,g, M/H) are adopted from \citet{Mathur2017} based on the latest \textit{Kepler} data release~25. \citet{Reinhold_sub} further used Gaia DR2 data to construct a Hertzsprung-Russell diagram (HRD), and select only stars confined between two isochrones: the "lower" one with an age of 4\,Gyr and metallicity of -0.8~dex, and an "upper" isochrone with an age of 5\,Gyr and metallicity of 0.3~dex. Furthermore, we focus on stars with near-solar photometric variability. The solar variability changes over the activity cycle, reaching $\rm R_{var}$ values up to 0.18\% \citep{Reinhold_sub}. Thus, we restrict our sample to stars with $\rm R_{var} < 0.18\%$. These selection criteria yield 911 stars in total. 
% Furthermore, we avoid contamination of our sample by subgiants. For that, we cross-matched the stars in our samples with the Gaia catalog using a 4~arcsec search radius\footnote{This search radius corresponds to the pixel scale of the Kepler telescope; see https://gaia-kepler.fun/}. 

Figure~\ref{fig:0100} shows the metallicity distribution in our sample.  The distribution has a pronounced peak at around M/H = -0.1. However, the uncertainties are quite large (up to $\pm 0.3 \rm dex  $).
To account for measurement uncertainties in the metallicity values, we describe the stars by a density distribution function, rather then using discrete values. For each star we used a normalised Gaussian distribution function, where the mean is set to the star's metallicity value and for the standard deviation we take the measurement uncertainty. The sum over all 911 metallicity distribution functions is shown by the red curve in Fig.~\ref{fig:0100}. 

To investigate a link between metallicity and the detectability of rotational periods, we further distinguish between stars with detected rotational periods (Table~1 in \citealt{McQuillan2014}), and without a significant period detection (Table~2 in \citealt{McQuillan2014}).  Hereafter, these two samples are referred to as the periodic and the non-periodic samples, respectively.  For stars in the periodic sample, we  restrict the rotation periods to the range 24--30~days.  Unfortunately, \citet{McQuillan2014} found rotation periods for only 10~stars in our sample. This number of stars is not sufficient to get a meaningful metallicity distribution of stars with known rotation periods.
%using the above selection criteria. 
Thus, we searched for potentially undetected periods for all stars in Table~2 of \citet{McQuillan2014} which satisfy our selection criteria. Analysing the full 4-yr \textit{Kepler} time series using Lomb-Scargle periodograms and auto-correlation functions, we were able to derive rotation periods for additional 87~stars (for details see Sect.~\ref{Prot_detection} and Table~\ref{parameters} in the appendix). Thus, the above constraints yield 97~stars with determined rotation periods, and  814~stars with unknown rotation period.

Figure~\ref{fig:0101} shows the detection fraction as a function of metallicity. For that we used the ratio of the metallicity distributions of stars with known rotation periods, $\rm n_{per} \rm (M/H)$, to the total number of stars (with known and unknown rotational periods), $\rm n_{all} \rm (M/H)$.
Our null hypothesis is that there is no effect of metallicity on the detection fraction, such that we expect a constant value (97/911), which is indicated by black dots in Fig.~\ref{fig:0101}. % For a reference value, we assume that there is no effect of metallicity and stars are equally distributed. 
%Consequently, we calculated the expected ratio as the ratio of stars with detected rotation periods to all stars  in the sample. 

We evaluated the statistical significance of the calculated detection fractions at seven metallicity values. For that, we randomly chose  97 stars from the total sample.  We repeated this 10,000 times, and calculated the mean value of the detection fraction at  seven metallicity values between $\rm M/H = -0.3$ and $\rm M/H = 0.3$. The mean values are identical with the expected ratio. Subsequently, the standard deviation from the expected ratio is obtained from the 10,000 realisations, and indicated by the error bars. 

Remarkably,  solar-like stars in the \textit{Kepler} sample show a similar trend as found for the  generated light curves for different metallicities (see Sect.~\ref{subsec:Modelling}). The detection fraction shows a minimum around solar metallicity,  and increases towards lower and higher metallicities.  In the range $-0.15 < \rm M/H < 0.15 $ the obtained fractions are below the standard deviation, i.e.~significant. 
%Note,  that the majority of the stars in our sample (over 65\%) are in the range of $ \rm -0.25 \le M/H \le 0.05$.
At the same time,   $\rm M/H < -0.15$ and $\rm M/H > 0.15$ the detection fractions remain within the error bars due to  large metallicity uncertainties and low number statistics.

%--- do not delete yet
%We note, the photometric variability $\rm R_{var}$ of these stars does not show a clear dependence on the metallicity (see lower panel in Fig.~\ref{fig:0102}).  Consequently, the observed decrease in the detection fraction with near-solar metallicity can only be brought about by the variability pattern in the light curves, and not by the amplitude of the variability itself. 

%%%%%%%%%%%%%%%%%%%%%%%%%%%%%%%%%%%%%%%%%%%%%%%%%%%%%%%%%%%%%%%%%%%%%%%%%%%%%%%%%%%%%%%%%%%%%%%%%%%%%%%%%%%%%%%%%
% Conclusion
%%%%%%%%%%%%%%%%%%%%%%%%%%%%%%%%%%%%%%%%%%%%%%%%%%%%%%%%%%%%%%%%%%%%%%%%%%%%%%%%%%%%%%%%%%%%%%%%%%%%%%%%%%%%%%%%%
%
\section{Conclusions}

The \textit{Kepler} field of view was selected such as to contain a large fraction of solar-like stars. Focusing on stars in the effective temperature range  of 5600K -- 5900K, 
it is challenging to determine their rotational periods \citep[see Fig.~14 in][]{vanSaders2019}. Moreover, \citet{Reinhold_sub} showed that the majority of the solar-like stars for which rotational periods could not be determined exhibit near-solar photometric variability, $\rm R_{var}$. Here, we provided one possible explanation for a decreased  detection fraction for these stars. 
%Here, we provided one possible explanation for a decreased  detection fraction for stars with near-solar rotation periods. }}

We investigated the compensation of the facular- and spot contributions to the rotational signal in stellar photometric light curves. This compensation hampers the detection of stellar rotational periods, and contributes to a decreased detection rate for  solar-like stars. Our results show that the detectability of rotational periods in stars with solar-like photometric variability crucially depends on metallicity. Hence,  metallicity is one additional important parameter to consider when investigating solar-like stars, and more accurate measurements of metallicity are needed.

\begin{acknowledgements}
This work has received funding from the European Research Council (ERC) under the European Union's Horizon 2020 research and innovation programme (grant agreement No. 715947). This work has been partially supported by the BK21 plus programme through the National Research Foundation (NRF) funded by the Ministry of Education of Korea.  We would like to thank the International Space Science Institute, Bern, for their support of science team 446 and the resulting helpful discussions. We would like to thank Jennifer van Saders for helpful discussion. 
\end{acknowledgements}

\bibliographystyle{aa} 
\bibliography{bib} 

%%%%%%%%%%%%%%%%%%%%%%%%%%%%%%%%%%%%%%%%%%%%%%%%%%%%%%%%%%%%%%%%%%%%%%%%%%%%%%%%%%%%%%%%%%%%%%%%%%%%%%%%%%%%%%%%%
% Appendix
%%%%%%%%%%%%%%%%%%%%%%%%%%%%%%%%%%%%%%%%%%%%%%%%%%%%%%%%%%%%%%%%%%%%%%%%%%%%%%%%%%%%%%%%%%%%%%%%%%%%%%%%%%%%%%%%%
%
\begin{appendix}

\section{\textit{Kepler} rotation periods}
\label{Prot_detection}
\begin{figure}
    \resizebox{\hsize}{!}{\includegraphics{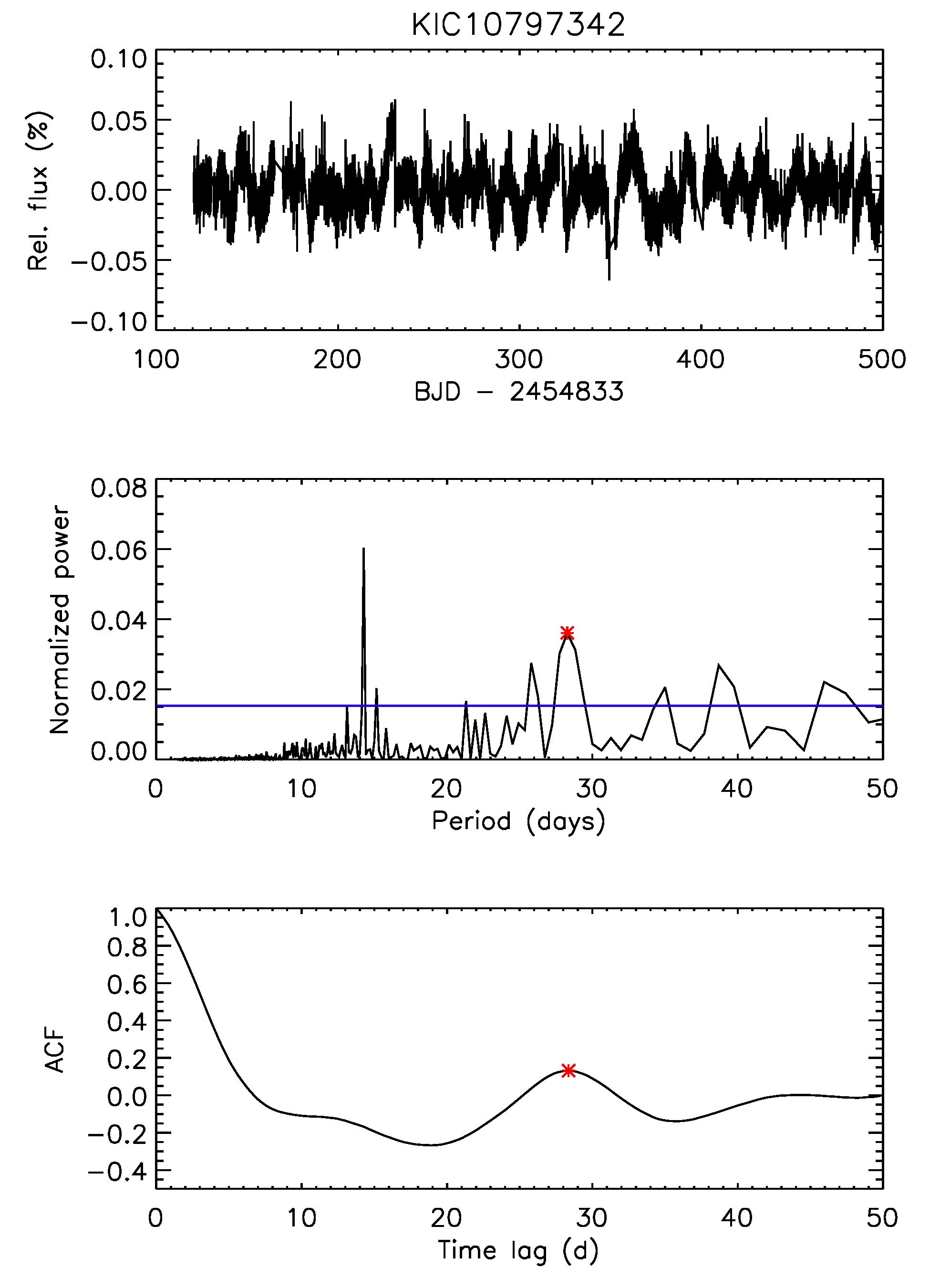}}
    \caption{Period detection for the star KIC\,10797342. Top panel: Part of the full 4-yr light curve showing photometric variability. Middle panel: Lomb-Scargle periodogram showing several peaks above the $5\sigma$ threshold (blue line). Bottom panel: auto-correlation function showing one pronounced peak indicated by the red asterisk.}
    \label{fig:example_lc1}
\end{figure}

 We analysed stars from Table~2 in \citet{McQuillan2014} satisfying our selection criteria (i.e. the non-periodic sample) to search for potentially undetected rotation periods. 
The \textit{Kepler} data used in this study were reduced with the so-called Presearch Data Conditioning (PDC) pipeline (version 8.0-9.2). We appended all available \textit{Kepler} quarters by dividing each light curve by its median and subtracting unity. Then we rebinned the data from 30-minute to 6-hour cadences, and analysed the full 4-year time series using Lomb-Scargle periodograms and auto-correlation functions. The Lomb-Scargle periodogram return peaks of different power on a period grid. The peak height is equivalent to the goodness of sine fit to the data, i.e. the more sinusoidal the signal, the higher the peak at this period. At the same time, the auto-correlation function (ACF) searches for self-similarity of the light curve signal, independently of the shape of the signal, and returns a peak at the best time lag (i.e. the ACF period). Since the time series of these slowly-rotating solar-like stars are usually non-sinusoidal, we use the ACF period as initial period guess. For both the periodogram and the ACF, we searched for cases where both methods returned \textit{consistent} peaks, with periods up to 50~days. The term \textit{consistent} covers different cases: a) the periodogram and the ACF both show a pronounced peak at roughly the same period, i.e. the periods may differ by up to five days. This case is considered the easiest one, and we omit showing an example here. Case b) the ACF shows a pronounced peak and the periodogram also shows a peak at this period, which still lies above the $5\sigma$ threshold (blue line). The highest periodogram peak lies at the half period, i.e. the first harmonic of the rotation period. Such a case is shown in Fig.~\ref{fig:example_lc1}, and for this star we report a rotation period $\rm P_{rot}=28.3$~days. Case c) the ACF shows a peak, and the periodogram shows many peaks of different height close  to the ACF period. The outer envelope of these peaks has a similar shape as the ACF peak.
The various cases described above show that it is non-trivial to assign a significance to a certain peak. As an additional test, each light curve was inspected by eye to search for the detected periodicity. Cases where it was not clear whether the detected period is the correct rotation period or only the half period (i.e. the first harmonic) were discarded. We distinguish three different cases for a non-detection.  First, when no peak above the $5\sigma$ threshold is detected, or second, when the periodogram shows several peaks of the same height, and it is not clear which of them is the correct rotation period. The third case is when  the periodogram shows a significant peak, while the ACF shows no peak, or a peak at a period that is not a harmonic of the peak detected by the periodogram.
\clearpage
\onecolumn
\begin{longtable}{cccccccc}
% \hline\hline
\toprule
KIC & $T_{\rm eff}$ & $\log\,g$ & M/H & $P_{\rm rot}$ & $R_{\rm var}$ & Kp & $\rm M_G$ \\
 & (K) & (dex) & (dex) & (days) & (\%) & (mag) & (mag) \\
\midrule
8145579 & 5788 & 4.43 & 0.00 & 25.04 & 0.18 & 14.63 & 4.91 \\
8166764 & 5662 & 4.56 & -0.12 & 28.45 & 0.17 & 14.86 & 5.06 \\
9653576 & 5736 & 4.58 & -0.34 & 24.37 & 0.13 & 14.28 & 5.00 \\
9725697 & 5883 & 4.40 & 0.34 & 28.25 & 0.12 & 14.21 & 4.36 \\
10203041 & 5758 & 4.45 & 0.10 & 27.40 & 0.17 & 14.71 & 5.36 \\
10331397 & 5833 & 4.55 & -0.24 & 26.56 & 0.16 & 14.25 & 4.66 \\
10449768 & 5612 & 4.57 & -0.12 & 28.09 & 0.13 & 14.05 & 5.34 \\
10876483 & 5651 & 4.57 & -0.16 & 25.29 & 0.17 & 14.89 & 5.53 \\
10979213 & 5706 & 4.30 & -0.14 & 28.58 & 0.17 & 13.52 & 5.43 \\
11097788 & 5815 & 4.52 & -0.26 & 24.73 & 0.16 & 14.01 & 5.26 \\
\midrule
\multicolumn{8}{c}{New periods} \\
\midrule
3526705 & 5612 & 4.57 & -0.12 & 28.75 & 0.12 & 13.86 & 4.82 \\
4246871 & 5880 & 4.45 & -0.28 & 29.99 & 0.09 & 13.21 & 4.76 \\
4458465 & 5814 & 4.52 & -0.22 & 29.78 & 0.14 & 14.76 & 4.78 \\
5017695 & 5835 & 4.51 & -0.02 & 28.45 & 0.15 & 14.84 & 4.62 \\
5121989 & 5881 & 4.50 & -0.20 & 28.61 & 0.15 & 14.00 & 4.35 \\
5272366 & 5692 & 4.55 & -0.04 & 28.00 & 0.13 & 13.94 & 4.88 \\
5444665 & 5706 & 4.48 & -0.28 & 27.70 & 0.07 & 12.99 & 5.02 \\
5528572 & 5750 & 4.33 & -0.18 & 27.50 & 0.09 & 14.61 & 4.59 \\
5608120 & 5682 & 4.57 & -0.28 & 28.00 & 0.11 & 14.29 & 5.38 \\
5617171 & 5687 & 4.41 & -0.14 & 29.80 & 0.09 & 14.43 & 4.99 \\
5697935 & 5862 & 4.55 & -0.28 & 29.18 & 0.18 & 14.71 & 4.32 \\
5737659 & 5662 & 4.42 & -0.12 & 27.80 & 0.10 & 14.12 & 4.85 \\
5773497 & 5811 & 4.30 & -0.24 & 27.43 & 0.09 & 14.80 & 5.18 \\
5809639 & 5760 & 4.56 & -0.26 & 27.42 & 0.13 & 14.31 & 4.72 \\
5893039 & 5672 & 4.53 & -0.04 & 24.10 & 0.10 & 12.92 & 4.97 \\
5955074 & 5791 & 4.51 & 0.21 & 26.76 & 0.09 & 14.21 & 4.49 \\
6025356 & 5704 & 4.57 & -0.30 & 26.91 & 0.14 & 14.52 & 5.00 \\
6049640 & 5628 & 4.57 & -0.22 & 27.80 & 0.17 & 14.79 & 5.22 \\
6109802 & 5808 & 4.39 & -0.14 & 28.83 & 0.04 & 13.59 & 5.27 \\
6358993 & 5825 & 4.56 & -0.30 & 27.50 & 0.07 & 13.53 & 5.27 \\
6542923 & 5799 & 4.55 & -0.16 & 24.90 & 0.15 & 14.44 & 4.84 \\
6604287 & 5684 & 4.31 & 0.14 & 27.53 & 0.10 & 14.35 & 5.12 \\
6613586 & 5802 & 4.44 & 0.00 & 25.00 & 0.12 & 13.60 & 4.72 \\
6676588 & 5809 & 4.56 & -0.26 & 29.18 & 0.05 & 14.45 & 4.66 \\
6931980 & 5631 & 4.56 & -0.06 & 29.17 & 0.09 & 14.89 & 5.01 \\
7293364 & 5818 & 4.34 & -0.26 & 25.35 & 0.12 & 13.12 & 4.97 \\
7335233 & 5822 & 4.56 & -0.28 & 26.06 & 0.09 & 13.93 & 4.65 \\
7335622 & 5810 & 4.48 & 0.00 & 28.05 & 0.10 & 14.63 & 4.68 \\
7586250 & 5755 & 4.56 & -0.18 & 29.00 & 0.07 & 14.42 & 5.32 \\
7590688 & 5647 & 4.52 & -0.16 & 24.10 & 0.14 & 10.92 & 4.96 \\
7757374 & 5861 & 4.55 & -0.20 & 28.35 & 0.13 & 14.84 & 4.18 \\
7770347 & 5611 & 4.22 & -0.06 & 26.47 & 0.15 & 14.10 & 5.28 \\
7875239 & 5812 & 4.51 & 0.02 & 29.17 & 0.08 & 14.78 & 5.30 \\
7885022 & 5841 & 4.52 & 0.07 & 26.23 & 0.08 & 13.86 & 4.53 \\
7886445 & 5727 & 4.48 & -0.22 & 24.86 & 0.08 & 13.34 & 5.00 \\
7947993 & 5899 & 4.37 & -0.08 & 28.09 & 0.15 & 14.90 & 4.17 \\
7977013 & 5635 & 4.51 & -0.18 & 26.44 & 0.08 & 14.91 & 5.19 \\
8013186 & 5665 & 4.57 & -0.20 & 28.09 & 0.16 & 14.72 & 5.37 \\
8046960 & 5876 & 4.29 & -0.30 & 27.29 & 0.05 & 13.85 & 4.13 \\
8076706 & 5823 & 4.28 & 0.14 & 27.02 & 0.12 & 11.10 & 4.51 \\
8107782 & 5738 & 4.28 & -0.04 & 27.82 & 0.15 & 13.66 & 5.21 \\
8212496 & 5790 & 4.53 & -0.02 & 24.82 & 0.18 & 14.14 & 4.50 \\
8222395 & 5896 & 4.52 & -0.10 & 24.23 & 0.07 & 13.50 & 4.73 \\
8478298 & 5627 & 4.52 & -0.22 & 28.36 & 0.17 & 13.71 & 5.30 \\
8629701 & 5809 & 4.56 & -0.26 & 26.73 & 0.06 & 14.54 & 4.40 \\
8702747 & 5876 & 4.30 & 0.21 & 24.87 & 0.06 & 13.64 & 4.35 \\
8812874 & 5805 & 4.47 & 0.30 & 28.22 & 0.08 & 14.65 & 4.35 \\
8881423 & 5876 & 4.56 & -0.32 & 25.28 & 0.15 & 14.31 & 5.00 \\
9040864 & 5774 & 4.30 & -0.12 & 27.81 & 0.14 & 14.69 & 5.29 \\
9204238 & 5853 & 4.53 & -0.10 & 27.96 & 0.06 & 13.86 & 4.96 \\
9221980 & 5830 & 4.42 & 0.24 & 28.90 & 0.08 & 14.40 & 4.77 \\
9326963 & 5666 & 4.40 & 0.18 & 26.33 & 0.11 & 14.50 & 5.18 \\
9409704 & 5653 & 4.54 & 0.00 & 29.04 & 0.17 & 14.28 & 4.98 \\
9451654 & 5696 & 4.52 & -0.14 & 29.44 & 0.14 & 14.74 & 5.21 \\
9457551 & 5609 & 4.50 & 0.04 & 28.93 & 0.07 & 13.37 & 5.14 \\
9508956 & 5895 & 4.28 & -0.32 & 24.29 & 0.16 & 12.97 & 4.48 \\
9518310 & 5737 & 4.40 & -0.16 & 27.95 & 0.13 & 14.55 & 4.56 \\
9706784 & 5657 & 4.54 & 0.07 & 29.99 & 0.14 & 14.24 & 5.24 \\
9773333 & 5864 & 4.47 & 0.24 & 28.84 & 0.13 & 13.53 & 4.44 \\
9821774 & 5876 & 4.47 & -0.04 & 25.94 & 0.07 & 13.38 & 4.51 \\
9835972 & 5847 & 4.48 & -0.14 & 29.02 & 0.11 & 13.29 & 4.80 \\
9843743 & 5804 & 4.55 & -0.18 & 27.47 & 0.05 & 14.20 & 4.50 \\
9904930 & 5786 & 4.22 & -0.12 & 24.38 & 0.07 & 12.37 & 4.40 \\
10002413 & 5704 & 4.43 & -0.20 & 28.89 & 0.13 & 14.49 & 5.30 \\
10002517 & 5759 & 4.50 & -0.20 & 28.46 & 0.11 & 13.08 & 4.98 \\
10018842 & 5803 & 4.50 & -0.22 & 26.30 & 0.07 & 12.40 & 5.11 \\
10024906 & 5838 & 4.53 & -0.08 & 29.09 & 0.14 & 14.89 & 4.94 \\
10026901 & 5692 & 4.55 & -0.04 & 25.71 & 0.10 & 14.81 & 4.90 \\
10079397 & 5799 & 4.54 & -0.10 & 28.36 & 0.15 & 14.72 & 4.90 \\
10122684 & 5699 & 4.34 & -0.22 & 28.09 & 0.08 & 14.09 & 4.99 \\
10155726 & 5767 & 4.56 & -0.22 & 24.86 & 0.04 & 13.15 & 4.96 \\
10198779 & 5762 & 4.48 & 0.16 & 28.63 & 0.11 & 14.94 & 4.89 \\
10330480 & 5795 & 4.46 & -0.26 & 26.98 & 0.07 & 14.87 & 5.18 \\
10419040 & 5679 & 4.45 & 0.28 & 29.43 & 0.06 & 14.39 & 5.10 \\
10552162 & 5819 & 4.57 & -0.32 & 27.74 & 0.08 & 14.27 & 5.04 \\
10722289 & 5831 & 4.54 & -0.16 & 24.32 & 0.12 & 13.29 & 5.26 \\
10794926 & 5806 & 4.50 & -0.02 & 27.02 & 0.16 & 12.95 & 4.98 \\
10797342 & 5783 & 4.49 & -0.32 & 28.36 & 0.04 & 12.63 & 4.40 \\
10849918 & 5815 & 4.53 & -0.30 & 29.41 & 0.05 & 13.60 & 5.14 \\
10924986 & 5755 & 4.50 & -0.04 & 28.09 & 0.06 & 13.65 & 5.17 \\
11019371 & 5662 & 4.57 & -0.26 & 28.77 & 0.10 & 14.90 & 5.37 \\
11038074 & 5864 & 4.36 & -0.26 & 27.02 & 0.11 & 13.29 & 5.00 \\
11136629 & 5778 & 4.54 & -0.30 & 27.28 & 0.05 & 14.05 & 4.74 \\
11181711 & 5741 & 4.43 & -0.14 & 29.46 & 0.14 & 14.72 & 4.72 \\
11860446 & 5785 & 4.57 & -0.34 & 26.21 & 0.09 & 13.56 & 5.17 \\
12218254 & 5655 & 4.58 & -0.30 & 26.20 & 0.12 & 14.32 & 5.35 \\
12505697 & 5864 & 4.35 & 0.18 & 29.81 & 0.09 & 13.48 & 4.95 \\
\bottomrule
\caption{Stellar fundamental parameters and rotation periods of the 97~periodic stars in our sample. In the last two columns the apparent Kepler magnitude Kp and the absolute Gaia magnitude $\rm M_G$ are given.}
\label{parameters}
\end{longtable}
\clearpage
\twocolumn

\section{Centre-to-limb variation}
\label{app:sec:02}

To investigate the behaviour of the power spectrum for the facular contribution we calculate the centre-to-limb variation, CLV, for the facular contrast. 
The facular  power spectra for different metallicity show  differently strong pronounced harmonics (see Fig~\ref{fig:0203}). The strength of the harmonics is affected by the shape of the light curve caused by a single faculae transiting   and thus depends on the centre-to-limb variations of facular brightness.

\begin{figure}
    \centering
    \resizebox{\hsize}{!}{\includegraphics{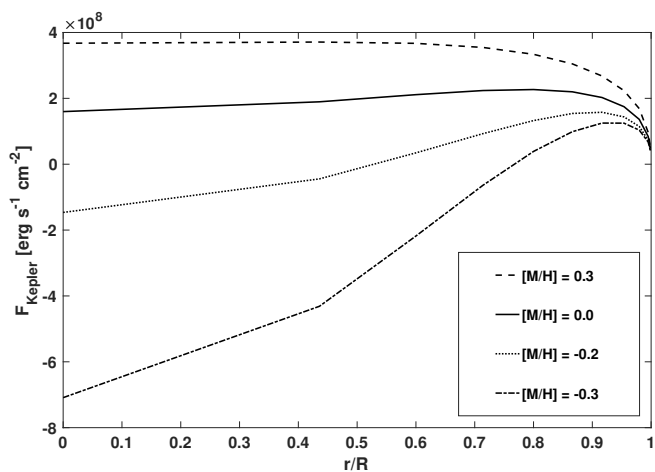}}
    \caption{Centre-to-limb variation of facular flux difference. Here r/R is the normalised radial distance from the centre of the stellar disc. Flux differences for different metallicities are shown in the \textit{Kepler} pass band.}
    \label{fig:app02}
\end{figure}

Figure~\ref{fig:app02} shows the CLV of the facular contrast in the \textit{Kepler} pass band for the solar case, and the three different metallicity values $\rm {M/H}=\{ -0.3,  -0.2,  0.3\}$. Most cases correspond to  the considered cases in Subsection~\ref{subsec:Modelling}.  The facular contrast is multiplied by the corresponding $\mu$, which is the cosine of the angle between the observer’s direction and the local stellar radius, to account for the foreshortening effect.

For the solar case, the facular contrast increases from the disc centre outwards.  This increase continues almost to the edge of the disc, before it drops again. Such an CLV will result in a doubled peaked feature transition curve, which leads to several pronounced harmonics in the power spectrum (see Fig.~\ref{fig:0203}a). While for a feature transition all harmonics are present in the power spectra, it will depend on the shape of the feature transition curve, which of them are more pronounced. For a more detailed discussion see \citet{paperI_Eliana_Sasha}.

On the contrary, for the case with greater metallicity ($\rm M/H = 0.3$), the facular contrast in the middle of the disc is almost constant, while a steep drop starts at $r/R \approx 0.7$. This results in an almost smooth peak in the feature transition curve, and thus less pronounced harmonics (see Fig.~\ref{fig:0203}e).
For the cases with $\rm M/H = -0.2$ and $-0.3$, a similar dip in the contrast appears at the edge and disc centre as in the solar case, which also leads to a double peaked feature transition curve. In addition, the contrast becomes negative towards the disc centre.  The facula  is only brighter at the limb, but becomes dark at the disc centre.

% \clearpage

\end{appendix}
\end{document}